# Precision Studies of Dark Energy with LSST[1]

J. Anthony Tyson and the LSST Collaboration

*Department of Physics, University of California, Davis, CA 95616*

**Abstract.** Starting around 2013, data from the Large Synoptic Survey Telescope (LSST) will be analyzed for a wide range of phenomena. By separately tracing the development of mass structure and rate of expansion of the universe, these data will address the physics of dark matter and dark energy, the possible existence of modified gravity on large scales, large extra dimensions, the neutrino mass, and possible self interaction of dark matter particles

## A WINDOW ONTO NEW PHYSICS

Dark matter and dark energy offer unique windows onto new physics, bridging particle physics and cosmology. While evidence for physics beyond the standard model has emerged from a variety of measurements -- solar and atmospheric neutrinos to high energy experiments -- striking evidence comes from the existence of the dark sector of our universe at late times and low energies. Our universe appears to be composed mainly of unknown forms of non-luminous mass-energy: 96 percent of the mass-energy is "dark". Over the last decade multiple measurements have led to a "standard model" of cosmology containing two mysterious new components: non-baryonic dark matter and dark energy. Non-baryonic dark matter implies the existence of a totally new sector of particle physics, which dominates the matter inventory of the universe.

### Dark Energy

Two independent lines of evidence point to accelerated expansion of our universe at late times. Within Einstein's theory, the underlying cause is a component of energy with large negative pressure referred to as dark energy. Dark energy accounts for 2/3 of the mass-energy in the universe, and is outside the current standard model. Indeed, a direct vacuum energy calculation yields an estimate 120 orders of magnitude larger than the value observed. Even in the context of supersymmetry this is only reduced to a 60 order of magnitude discrepancy.

Dark energy affects the cosmic history of the Hubble expansion $H(z)$ as well as the cosmic history of mass clustering. If combined, different types of probes of the expansion history and structure history can lead to percent level precision in dark energy parameters. This is because each probe depends on the other cosmological parameters or errors in different ways. These probes range from cosmic shear, baryon

---

acoustic oscillations, supernovae, and cluster counting -- all as a function of redshift $z$. Using the CMB as normalization, the combination of these probes will yield the needed precision to distinguish between models of dark energy. What is required is a facility which can undertake all of these probes with deep data over wide area with cross checks to control systematic error.

Current data merely constrain the existence of the effect but little else. Of particular interest is the dynamical behavior of dark energy, i.e. how it behaves with cosmic time or with redshift. It is common to characterize this evolution in terms of an "equation of state parameter": $w \equiv p/\rho$, where $p$ is the pressure, and $\rho$ is the energy density. For a pure cosmological constant, $w = -1$, and is constant in time. If the dark energy is associated with a new scalar field, one might expect $w$ to be a function of the scale factor of the universe, usually represented by $a$, where $a = 1$ in the current epoch ($z = 0$), and $a = 0$ at the time of the Big Bang. A simple parametrization for the cosmic pressure to energy density ratio $w(z)$ is then $w(a) = p/\rho = w_0 + w_a (1-a)$. Current data only constrain $w_0$ to be consistent with $-1$ to within $10 - 20\%$ (depending on assumptions), and place no meaningful constraints on $w_a$.

## LSST

Three studies conducted by the National Research Council have endorsed the construction of a wide-field telescope that will rapidly survey the entire visible sky to extremely faint limiting magnitudes. Advances in microelectronics, large optics fabrication, and computer hardware and software now make it possible. The Large Synoptic Survey Telescope (LSST) system will obtain sequential images of the entire observable sky every week. These images can be co-added to provide unprecedented depth and area coverage, yielding precision probes of dark energy. The positions in 3-D, shapes, colors, and brightness of distant galaxies provide precision constraints on the physics of dark energy and dark matter. About 200,000 distant galaxies per square degree distributed up to redshift 3 may be resolved from ground-based optical telescopes located in excellent sites because the galaxy's characteristic angular scales [1] are larger than the point-spread due to atmospheric turbulence. While the atmosphere introduces correlations in point-spread ellipticity on small angular scales, recent data suggests that this systematic error on distant galaxy ellipticites may be reduced to insignificant levels by using foreground stars as calibrators [2].

Because of its unprecedented capabilities and its promise for discovery at the frontiers of astronomy and physics, the LSST has brought together scientists and engineers from many universities, Department of Energy laboratories, the National Optical Astronomy Observatory (NOAO), and the private sector. Together, this group has devised an 8.4-m telescope, a 3.2 Gigapixel camera system with a 10 square-degree field of view, and a suite of prototype image-processing pipelines. Like accelerators, the LSST will supply data to separate science programs simultaneously.

The LSST total effective system throughput or *étendue* (analogous to accelerator luminosity), $A\Omega = 318$ m$^2$ deg$^2$, is nearly two orders of magnitude larger than that of any existing facility. LSST will enable a wide variety of complementary scientific investigations, all utilizing a *common database*. The LSST design is shown in Fig. 1 and its étendue relative to other facilities in Fig. 2.

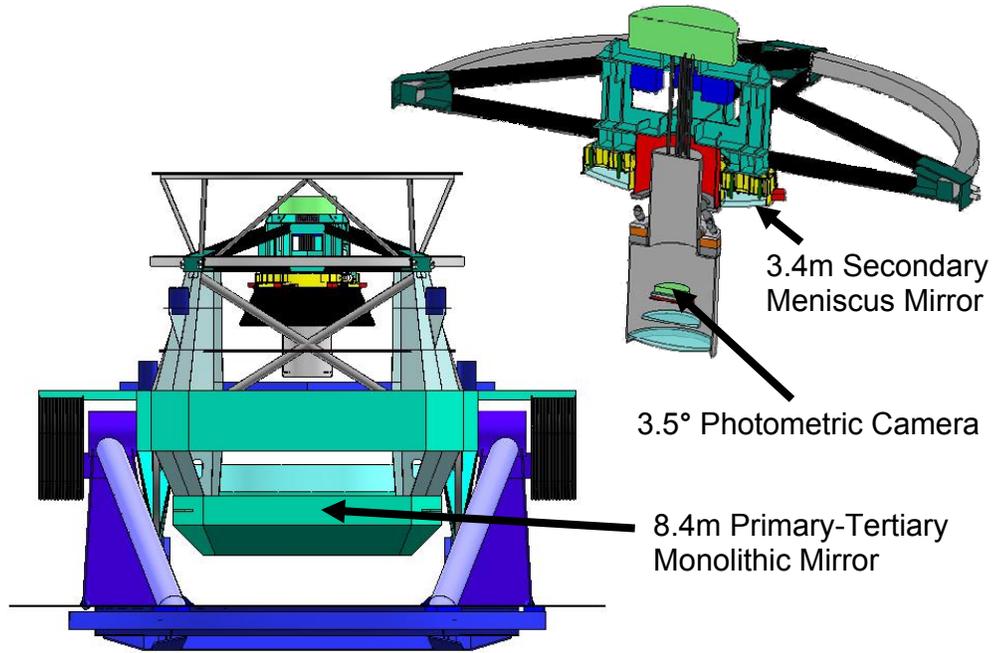

**FIGURE 1.** The 3 mirror LSST telescope design is shown on the left, and a cut-away of the top end on the right shows the additional optics in the 3 Gigapixel camera.

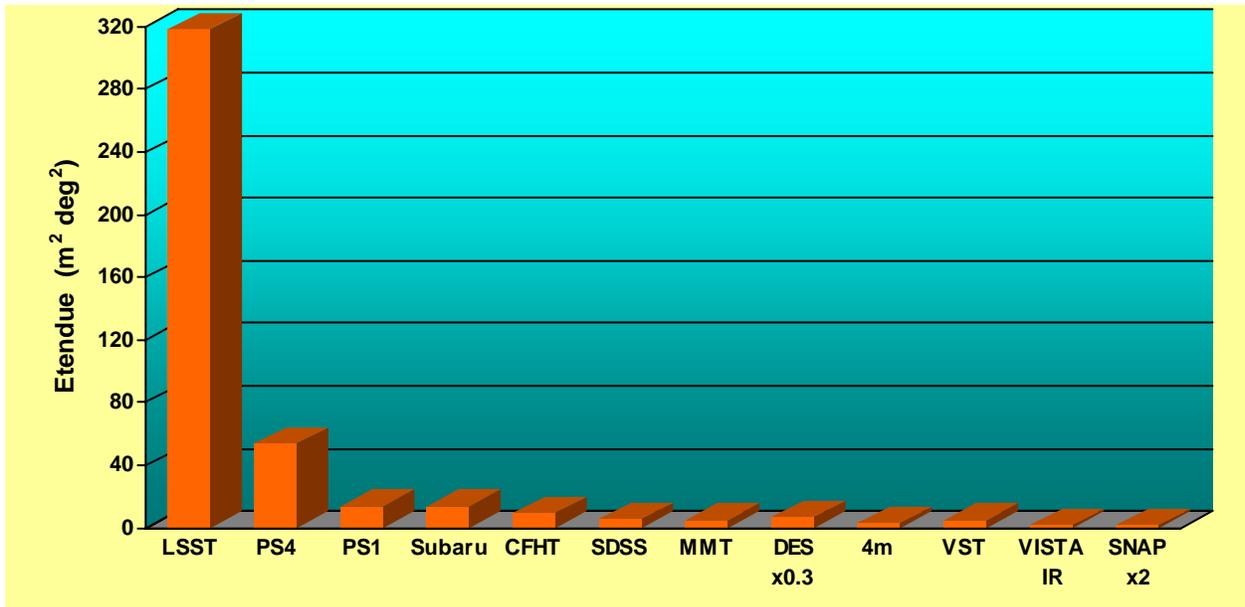

**FIGURE 2.** Survey power is proportional to the étendue (the $A\Omega$ product) of the telescope aperture and camera field of view in $m^2$ $deg^2$. This plot compares various imaging facilities, current and planned. The LSST will open up a qualitatively new regime in survey science. A unique result of a very high optical étendue is that many science programs can proceed in *parallel* with the same high quality data.

Of particular interest to particle physics, LSST will probe the physics of dark energy in multiple cross-calibrating ways, as well as measure the neutrino mass down to the 40 milli eV level. This will be done via the shapes, positions, and redshifts of billions of galaxies, plus a million supernovae. The physics of dark matter will be probed via the strong gravitational lenses LSST will discover. The physics that produces the observed accelerating cosmic expansion is a complete mystery. We are now capable of addressing the underlying dark energy physics by exploiting a diversity of precision cosmic probes:

- Weak lensing of galaxies vs redshift, which probes both the evolution of structure over cosmic time and dimensionless ratios of distances vs cosmic time, thereby setting multiple independent strong constraints on the nature of dark energy,

- Correlations of galaxies in vs cosmic epoch (Baryon Acoustic Oscillations) utilizes the "standard ruler" of the peak in the angular correlation of dark matter revealed in the temperature anisotropies in the cosmic microwave background.

- The redshift distribution of shear peaks due large clusters of dark matter (via weak gravitational lensing combined with the optical data) are an exponentially sensitive probe of the equation of state of dark energy.

- A million supernovae are a useful complementary technique for probing the recent cosmic era when dark energy becomes dominant.

By simultaneously measuring mass growth and curvature, LSST data can tell us whether the recent acceleration is due to dark energy or modified gravity [3,4]. Astrophysical observations are susceptible to sources of systematic error, so the LSST is being specifically designed and engineered to minimize and control systematics at a level ten times below the smallest signal. The diverse techniques listed above are complementary, removing degeneracies. Much of the power of the LSST comes from the fact that the measurements will be obtained from the same basic set of observations, using a powerful facility that is optimized for this purpose. A site has been chosen for the facility on Cerro Pachon in northern Chile.

## The Survey: Wide Fast Deep

LSST will collect photometric imaging data (position, flux, colors, shapes) on four billion galaxies in a survey covering 20,000 square degrees of the sky in six optical bands from 350nm to 1070nm. Flux levels fainter than 3E-5 photons cm$^{-2}$ sec$^{-1}$ will be reached at 10$\sigma$. Fitting to galaxy spectral templates vs redshift, these data yield redshifts to all these galaxies; multi-wavelength imaging photometry can be used to estimate the redshifts of every galaxy [5,6]. This selected sample of galaxies is distributed in redshift from z=0.1 to z~3. These source galaxies thus bracket well the epoch of maximal cosmological effects from dark energy at z~0.4. In addition, one million Type Ia supernovae will be frequently sampled with multi-band photometry and their redshift measured in a similar way [7].

# LSST Observational Windows on Dark Energy

Combined with cosmic microwave background data, four complementary investigations hold the key for probing dark energy: weak lensing cosmic shear [8,9], baryon acoustic oscillations (BAO) [10,11], supernovae [7], and cluster counting [12,13]. All four will be undertaken in parallel with LSST. For these four powerful cosmological probes it is necessary to calibrate photometric redshifts over the range in redshift 0<$z$<3 for faint galaxies and $z$<1 for supernovae. For this problem as well as control of shear systematics [14,15], the technology exists today to undertake the needed precursor programs. The prognosis looks good; initial experimental tests with new technology facilities on small areas of the sky show that it will be possible to control systematic errors to levels far below the dark energy signal. Here is a brief review of these four probes.

## *Weak Gravitational Lens Tomography*

Observables in weak gravitational lensing are in principle predictable ab initio given a cosmological model. Measurements are limited mainly by instrumental systematics rather than unknown astrophysics. LSST is uniquely capable of addressing the underlying physics by exploiting a diversity of cosmic probes. Background galaxies are mapped to new positions on the sky by intervening mass concentrations, shearing their images tangentially. The full 3-D cosmic mass distribution creates statistical correlations in galaxy shapes called "cosmic shear" [15]. Cosmic shear depends on the mass density field as a function of redshift, since the density field is what does the lensing of background galaxies, and it depends on the history of the expansion rate, since that determines the distance-redshift relation and therefore how length scales at a given redshift project into angular separations on the sky today. The recent report of the Dark Energy Task Force, names weak lensing (WL) as ``...likely to be the most powerful individual Stage-IV technique and also the most powerful component of a multi-technique program."

Weak lensing has high information content. A big advantage for lensing over CMB measurements is this ability to do 3-D tomography. If we know redshifts of the source galaxies, the mass distribution and cosmic geometry can be measured *as a function of redshift*. WL thus has sensitivity to the evolution of dark energy. With accurate calibration of the photometric redshifts, LSST's multi-color deep imaging survey will provide redshift information for distant galaxies to *z*=3. This additional information allows us to perform lensing tomography [8,9]. If the galaxies can be separated into *n* multiple redshift bins, then we can create *n* shear maps. The most interesting statistical properties of these maps are the two-point functions. These *n(n+1)/2* unique shear power spectra can be written as projections of the matter power spectrum along the line of sight out to some redshift. Jointly, these correlations contain enough information to break degeneracies and determine cosmological parameters, including dark energy parameters.

For WL and especially for BAO (below), it is necessary to accurately calibrate the photometric redshift measurements. A combination of spectroscopy and angular correlations of galaxies can produce the precision calibration [16].

## *Baryon Acoustic Oscillations*

Features in the matter power spectrum, caused by acoustic waves in the baryon-photon plasma prior to recombination [17], can serve as CMB-calibrated standard rulers for determining the angular-diameter distance *r(z)* and constraining dark energy. The scale is set by the sound horizon at decoupling and is reflected at later times in the 3-D distribution of galaxies. The lowest *k* peak in the series of baryon acoustic oscillations of the galaxy spectrum has been observed in spectroscopic redshift surveys of galaxies [10], but the samples are not large enough to provide sufficient precision for significant constraints on *r(z)*.

There are two advantages of a large photometric redshift survey such as the LSST: wide coverage (which reduces the sample variance error), and deep photometry (which leads to larger volume, more galaxies, and lower shot noise.) One may quantify this advantage with the effective survey volume $V_{eff} = \int d^3r \; n^2(r)P^2(k) [1+n(r)P(k)]^{-2}$, where *P(k)* is the galaxy distribution power spectrum, and *n(r)* is the galaxy number density. The challenges include photometric redshift errors, dust extinction, galaxy clustering bias, nonlinear redshift distortion, and nonlinear evolution. Despite the complexities, these uncertainties do not produce oscillating features in the power spectrum. However, the galaxy clustering bias (the relation between dark matter and galaxy density) must be calibrated; BAO data by itself does not provide this calibration. Cosmological parameter precision from BAO auto power spectra is also impacted by photometric redshift errors more than WL; however photo-z bias may be self calibrated by utilizing correlations between redshift bins -- cross power [18,19]. With galaxy bias data from a separate probe such as WL, and making use of BAO cross power spectra, it is possible to use a photometric redshift survey to measure the angular-diameter distance accurately.

## *The Cosmic Distribution of Projected Mass Peaks*

Another advantage of lensing over CMB measurements is that recent and/or small-scale fluctuations are *non-Gaussian*, i.e. there is information present beyond the power spectrum. For large numbers of source galaxies, three-point correlations of shear vs redshift become feasible and an LSST-scale survey is as powerful as the power spectrum in constraining dark energy parameters [9]. Weak lensing measurements on small scales are limited by the noise from the intrinsic ellipticities of the background galaxies. To average down this shot noise we need deep images to maximize the number of resolved galaxies per unit area on the sky.

The fractional area of shear peaks is exponentially sensitive to the dark energy equation of state. This is because these high shear regions are directly related to the dark matter structures which grow from the non-linear evolution of density fluctuations. This method can yield constraints on cosmological parameters that are comparable to those from the redshift distribution of galaxy cluster abundances, but without the problem of projections of physically unrelated mass overdensities [13]. Errors in the 3-D matter power spectrum *P(k)* increase the errors in cosmological parameters, so that *P(k)* has to be calibrated with accuracy of ~2% in order not to degrade the error in $w_0$ by more than 10%. While current accuracy in Monte Carlo calculations of $P(k)$ on these small scales is about 5-10%, the required level of 2% accuracy should be achievable in future numerical simulations.

*The LSST Supernova Samples*

Type Ia supernovae may be used as distance indicators if they are "standard candles." Current precision is ~7% in distance. LSST will find supernovae in two ways. The first is as a result of its normal operating mode providing frequent, all-sky coverage. The baseline observation strategy for the LSST survey will discover roughly one million Type Ia SNe per year. This sample will have a mean redshift of about 0.45 and extend to 0.7. The flux vs time (lightcurves) will typically have ~5 day sampling with less frequent sampling over all wavelength bands. Such a large sample of SNe over the sky, plus color redshifts and morphology of the host galaxies, will constrain the angular variation of dark energy parameters and provide an unprecedented opportunity to search for hints to the nature of the Type Ia progenitors and to search for "third parameters," especially by correlating with environmental properties.

A small fraction of LSST operational time will be devoted to producing a second supernova sample, from a "staring mode" deep search of a limited number of 10 square degree fields. Over a ten year run, 10-20 minutes per night spent staring in this mode will yield (with no rate evolution) ~100,000 supernovae with lightcurves of unprecedented detail, typically with more than 100 photometric points per supernova in five bands. This sample will have a mean redshift of 0.75 and extend to $z>1$. Such detailed lightcurves will allow fitting for photometric redshifts from the supernovae themselves; simulations show that such SN photometric redshifts have a typical error <0.01 in z. These can in turn be combined with host-galaxy photo-zs. Since the area on the sky is limited, multi-object spectroscopy can be obtained for many of these. The main advantage of such a large supernova sample is to investigate possible evolution in the "standard candle" by detecting correlations with other parameters.

## Probing the Physics of the Recent Acceleration

These various probes depend differently on the cosmological parameters. Thus combining multiple probes with the CMB data removes degeneracies. This is crucial if we want to determine all parameters from the data, without assumed priors. Due to its uniquely high étendue, the LSST survey will produce all four complementary probes of dark energy from the same survey data. Weak gravitational lensing is sensitive to angular diameter distance as a function of redshift and to trends in mass clustering with redshift. These same imaging data (with precision photometric redshifts) on four billion galaxies will track the baryon acoustic oscillations over cosmic time. Statistics of peaks in shear due to non-linear clustering provide another complementary probe. Finally, the tens of thousands of supernovae out to z~1 provide yet another cross check. When combined with the cosmic microwave background anisotropy data these tests form interlocking checks on cosmological models and the physics of dark energy.

The combination of BAO with WL is especially powerful [19]. Figure 3 shows an estimate of the precision in 2-parameter dark energy model space for the LSST survey by combining two of its four probes in a global fit to nine cosmological parameters.

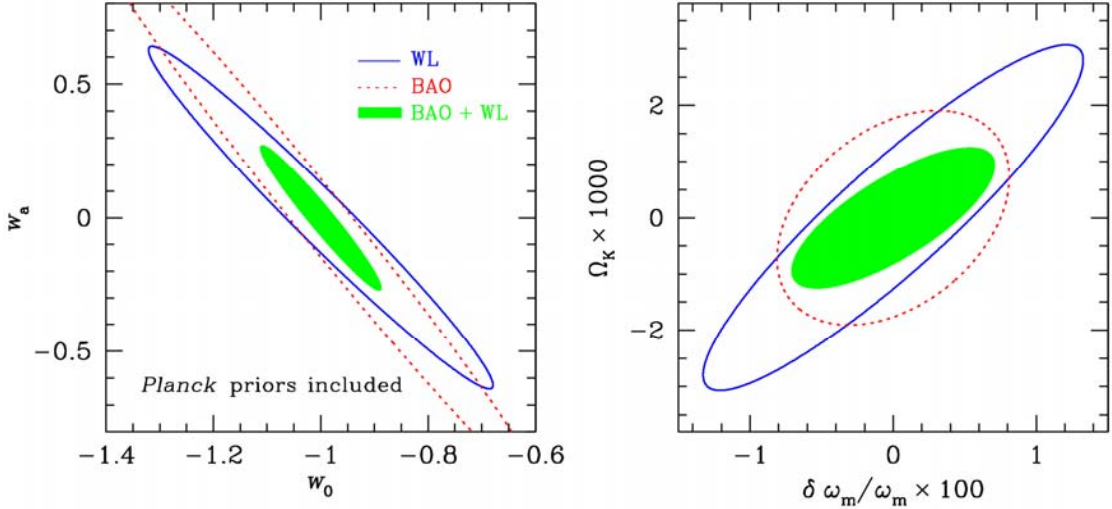

**FIGURE 3.** *Left panel*: Forecasts of LSST errors on the dark energy equation-of-state parameters $w_0$ and $w_a$ for BAO (dotted line), WL (solid line), and the two combined (shaded area). The constraints are marginalized over 9 other cosmological parameters including the curvature and over 120 parameters that model the linear galaxy clustering bias, photometric redshift bias, and rms photometric redshift error [19]. *Right panel*: Same as the left panel, but for the matter density $\omega_m$ and curvature term $\Omega_K$. Note that *Planck* alone will constrain $\omega_m$ to 1% (it does poorly on the curvature). Adding SN data improves the constraints only slightly. Adding shear peak statistics [13] and galaxy clustering power spectrum data can further increase the precision.

Adding cross-correlations with foreground galaxies increases the precision significantly [20]. Is our model too simple? Using galaxy $P(k)$ from these same data together with Planck data, LSST will usefully constrain six eigenmodes of the dark energy equation of state. These data provide us with an opportunity to discriminate between two very different explanations for the observed acceleration. This capability is due to the sensitivity of WL data not just to the history of the expansion rate, $H(z)$, but also to the rate of growth of the large-scale density field. Cosmic shear depends on the mass density field as a function of redshift, since the density field is what does the lensing of background galaxies, and it depends on the history of the expansion rate, since that determines the distance-redshift relation and therefore how length scales at a given redshift project into angular separations on the sky today.

One can model the observed cosmic shear power spectra as a function of redshift by varying both $g(z)$ and the distance-redshift relation, $D(z)$, as independent functions. The cosmic shear power spectra are capable of constraining both separately with the same data, leading to an important test. In Einstein gravity with dark energy, these two functions are not independent. With the dark energy properties adjusted to give the observed $D(z)$, a prediction can be made for $g(z)$. Theories of gravity with modified force laws on >Mpc scales will generally have different predictions for $g(z)$ [4].

## Dark matter and the total neutrino mass

Hints of solutions to these puzzles may come from the LHC and the ILC. We know however that there is a large signal in the cosmic geometry and the distribution of dark matter as a function of time. More refined cosmological measurements will be extremely important. There are more mysteries involving the "dark sector." For example, current data provide very weak constraints on how dark matter behaves dynamically, and how it interacts with itself or with baryonic matter. The clumpiness observed in the dark matter halos of galaxies appears less pronounced than expected for purely gravitational interactions. This needs confirmation with similar strong lensing studies of many more clusters. The LSST will discover several hundred thousand such clusters, hundreds of which will have the special alignment of a background galaxy giving rise to multiple images (strong lensing).

In addition to the strong lens probe of the dark matter distribution in cores of clusters, a hemisphere scale weak lens survey will precisely measure the power spectrum of dark matter fluctuations. The slope of this spectrum is sensitive to the sum of neutrino masses. If dark matter is completely cold (i.e. without a small "warm" component) then the effects of neutrino mass are unambiguous, and LSST should reach a sensitivity of 0.04 eV or better [21]. More details on these dark matter and dark energy probes may be found on the LSST website http://www.lsst.org

Is our universe an accident or are there dynamical mechanisms which generate the observed energy hierarchies of fundamental physics? Precision data on dark energy and dark matter may shed light on this issue.

## ACKNOWLEDGMENTS

LSST is a public-private partnership. The LSST research and development effort is funded in part by the National Science Foundation under Scientific Program Order No. 9 (AST-0551161) through Cooperative Agreement AST-0132798. Additional funding comes from private donations, in-kind support at Department of Energy laboratories and other LSSTC Institutional Members.

## REFERENCES

1. H. C. Ferguson, et al., *Astrophys. J. Lett.* **600,** 107-110 (2004)
2. D. Wittman, *Astrophys. J Lett.* **632**, 5-8 (2005)
3. J. Zhang, L. Hui, and A. Stebbins, *Astrophys. J.* **635**, 806-820 (2005)
4. L. Knox, Y.-S. Song, J.A.Tyson, *Phys. Rev. D.* **74**, 023512-023515 (2006)
5. A. J. Connolly, et al., *Astron. J.* **110**, 2655-2665 (1995)
6. N. Benítez, *Astrophys. J.*, **536**, 571–583 (2000)
7. B. J. Barris and J. L. Tonry, *Astropys. J.* **613**, 21 (2004)
8. W. Hu, *Astrophys. J. Lett.*, **522**, 21–24 (1999)
9. M. Takada, and B. Jain, *MNRAS* **348**, 897-915 (2004)
10. N. Padmanabhan, et al., *Astro-ph*/0605302 (2006)
11. H. Zhan, and L. Knox, *Astrophys. J.*, **644**, 663–670 (2006)
12. W. Hu, *Phys. Rev. D*, **67**, 081304 (2003)
13. S. Wang, Z. Haiman, M. May, and J. Kehayias, *Astro-ph*/0512513 (2005)
14. D. Huterer, M. Takada, G. Bernstein, and B. Jain, *MNRAS*, 366, 101-114 (2006)
15. D. Wittman, et al., *Nature* **405**, 143-149 (2000).
16. M. Schneider, L. Knox., H. Zhan, and A. Connolly, *Astro-ph*/0606098 (2006)
17. P.J.E. Peebles and J.T. Yu, *Astrophys. J.* **162**, 815-836 (1970)
18. L. Knox, Y.-S. Song, and H. Zhan, *Astro-ph*/0605536 (2006)
19. H. Zhan, *Astro-ph*/0605696 (2006)
20. W. Hu and B. Jain, *Phys. Rev. D* **70**, 043009-043016 (2004)
21. S. Hannestad, H. Tu and Y.Y. Wong, *JCAP* **06**, 025-050 (2006); *Astro-ph*/0603019